\documentclass[journal=jacsat,manuscript=article]{achemso}
\usepackage[version=3]{mhchem} % Formula subscripts using \ce{}
\usepackage[T1]{fontenc}       % Use modern font encodings

\author{Sven P. Rudin}
\email{srudin@lanl.gov}
\affiliation{Los Alamos National Laboratory, Los Alamos, New Mexico 87545, USA}

\title{A Density Functional Theory Calculations-Based Approach that
Predicts Layered Materials with Emergent Structures}

\begin{document}

\begin{abstract}
Recent work shows that density functional theory calculations
accurately describe materials exhibiting turbostratic disorder
between layers of incommensurate constituents.
These calculations approximate one of the constituents as a finite
island between continuous layers of the other constituent.
This island approximation allows equal treatment of all relative angles between the
interleaved layers, and it removes the requirement of a unit cell that
is commensurate with both layers.
The work presented here uses this approximation to search for new
layered materials
that exhibit turbostratic disorder and magnetic structure.
In a first application,
the approach finds the recently synthesized SnSe$_2$-VSe$_2$ system.
The approach predicts a stable Fe$_3$Ge-VSe$_2$ structure
and multiple stable structures that interleave Mn$_x$Se$_y$ layers with VSe$_2$ layers.
These Mn$_x$Se$_y$ layers exhibit emergent structures and stability.
\end{abstract}

The theoretical approach outlined here proposes to accelerate
the experimental discovery of
new layered materials.
Recent efforts in creating new materials that layer
known 2D materials reveal
properties and phenomena of both scientific and practical interest \cite{Geim:2013fk}.
Designing materials with a desired functionality becomes possible by
choosing the individual constituents with specific properties,
provided the interleaved constituents form a kinetically stable material.
Recent advances in high-throughput calculations of, e.g., thermoelectric properties
\cite{C5TC04339E}
enable a highly informed choice of possible constituents;
recent advances in synthesis methods enable their assembly into nanostructured
materials with high thermoelectric efficiency
\cite{moss1996, Chiritescu351, Karttunen2016338}.
Desired properties can also be strongly influenced by the interfaces between the layers,
in particular low thermal conductivity benefits from interleaving layers
with lattices that remain incommensurate \cite{lin2010}.
Allowing for incommensurate layers
significantly widens the pool of possible combinations in the search for new materials.

Confinement at the nanoscale also allows the constituents to stabilize in structures
that are neither stable as an isolated 2D form nor a slice of a stable bulk crystal structure.
Such emergent structures
can be discovered serendipitously, guided by
intuition and chemical similarity.
Theoretical guidance can focus experimental efforts on systems more likely to show success.
Density functional theory (DFT) \cite{lda2} calculations can provide such guidance,
testing all possible structures constructed from systematically
chosen combinations of chemical elements.
This type of approach is successfully applied to, e.g.,
calculating the defect energies and site preference of all possible dopants in
lithium battery materials
\cite{Miara2015}.
In the search for new layered materials, however,
the large number of possible structures and chemical element combinations
makes this kind of search costly. 
In addition, for each probed system
unit cells must be constructed that are commensurate with each pair of
potential structures of the layered constituents --
a challenging task when neither the layers' lattice constants nor their relative orientation
are known beforehand.

\begin{figure} %%%%%%%%%%%%%%%%%%%%%%%%%%%%%%%%%%%%%
\includegraphics[width=8.46cm]{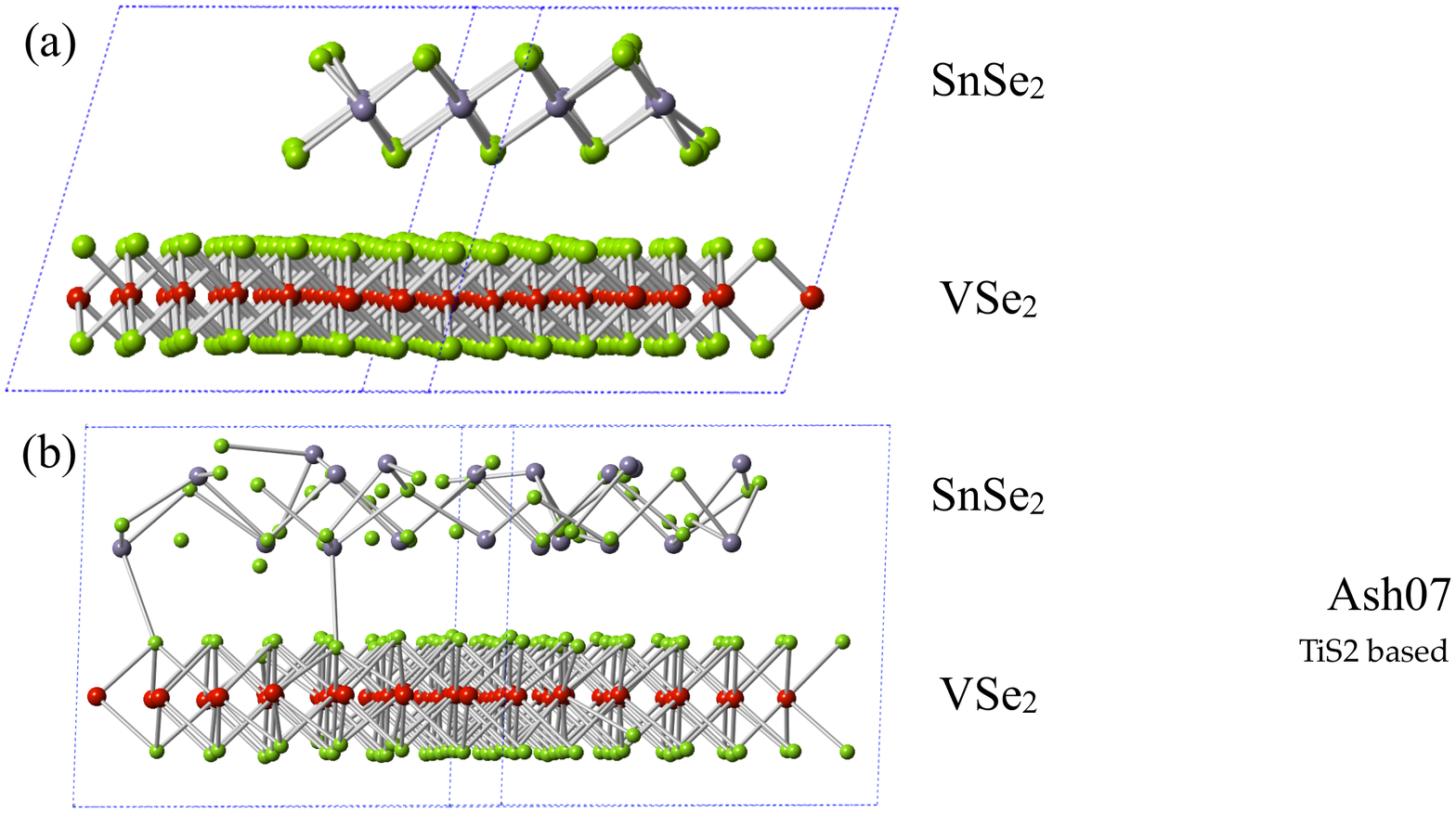}
\caption{\label{fig:SSV}
Side views of optimized unit cells containing a complete VSe$_2$ layer
and a SnSe$_2$ island structure.
(a) Optimization of a SnSe$_2$ island in the cadmium iodide structure
results in a promising case, in agreement with experiment \cite{kylepriv}.
(b) Starting from an initial SnSe$_2$ island in the molybdenum disilicide structure
results in an agglomeration of ions without much structure.
V, Se, and Sn ions are colored red, green, and grey, respectively.
Bars represent V-Se and Sn-Se bonds with lengths of up to 3 \AA.
}
\end{figure}

The approach outlined here proposes to accelerate
such theoretical searches by using unit cells that approximate one constituent as a finite
island between continuous layers of the other constituent.
This truncation,
used in Ref.~\citenum{PhysRevB.91.144203},
removes the requirement for an {\it a priori} constructed unit cell
that is commensurate with both layers.
The finite island possesses significantly more freedom to rearrange its ions
during DFT optimization, so multiple final structures become accessible from
a single starting configuration.
The search is further accelerated because each probed unit cell can be much smaller than
would be required for a complete, commensurate unit cell.
Edge effects arise, but experience shows that, after optimization, the core ions in the islands 
either distinctly resemble sections of a crystal structure or they appear randomly dispersed
(see Fig.\ \ref{fig:SSV}).
The former, promising systems can serve to guide synthesis efforts.
Details of these promising structures can also serve to construct unit cells in
which both constituents are continuous layers,
unit cells with which further theoretical study can test for interesting properties.
Experience with such unit cells shows their optimization takes on the
order of five times longer than the typical optimization of an unit cell with
one layer approximated as a finite island.
This differences translates into an acceleration of the
theoretical search for new materials that is at least five-fold,
which represents a lower limit because it includes only the effect of using
smaller unit cells and not the other points raised above.

The approximation with a finite island arises in applying DFT to understand
a material exhibiting turbostratic disorder, $[$(SnSe)$_{1+y}]_m$(VSe$_2$)$_n$
\cite{PhysRevB.91.144203}.
Layered materials with turbostratic disorder consist of parallel-layered constituents where the translation parallel to the layer and rotation normal to the layer are random on average
\cite{warren41, biscoe1942}.
Such disorder leads to ultralow thermal conductivity across the layers \cite{Chiritescu351}.
Layered materials with turbostratic disorder can also be designed for a tunable
superconducting transition temperature \cite{Grosse:2016aa}.
In $[$(SnSe)$_{1+y}]_m$(VSe$_2$)$_n$,
$n$ triatomic thick (001) slices of VSe$_2$ alternate with $m$
diatomic thick (001) slices of SnSe.
The two constituents exhibit a mismatch in formula units per area,
the misfit parameter $1+y$.
Approximating SnSe as a finite island between continuous layers of VSe$_2$
allows DFT calculations that treat equally all angles between the constituents.

Given the approximation's accurate description of an existing system,
this work uses it to predict systems
that could potentially be synthesized.
The investigated systems retain VSe$_2$ as one constituent and construct
the other from two period four elements
to probe the latter's viability when interleaved with VSe$_2$.
Specifically, three combinations are investigated:
(1) Sn$_x$Se$_y$, to test whether
the approach delivers the recently discovered SnSe$_2$-VSe$_2$ structure
\cite{kylepriv},
(2) Fe$_x$Ge$_y$ and
(3) Mn$_x$Se$_y$.
The inclusion of magnetic ions aims to probe for systems that exhibit
turbostratic disorder and magnetic structure.

\begin{table}
\begin{center}
\begin{tabular}{c c | c c c c  }
\multispan2{\hfil composition\hfil} & bulk structure type & slice & $1+y$ \cr
bulk & layer \cr
\hline
SnSe           &   SnSe                    & rock salt               & [001] & 1.19 \cr
SnSe$_2$  &  SnSe$_2$           & cadmium iodide & [001] & 1.36 \cr
\hline
Fe$_3$Ge  &  Fe$_2$Ge           & auricupride         & [001] & 1.39 \cr
\hline
Mn$_2$Se  & MnSe                    & iron stannide      & [001] & 1.52 \cr
                     & Mn$_3$Se$_2$  &                               &           & 1.40 \cr
Mn$_3$Se & Mn$_2$Se           & auricupride          & [001] &  1.38 \cr
MnSe$_2$  &  MnSe$_2$           & cadmium iodide & [001] & 1.24 \cr
MnSe$_2$  &  MnSe$_2$           & molybdenum disilicide & [100] & 2.6 \cr
\end{tabular}
\caption[]{
Details of predicted composition and structure of layers that form kinetically stable systems
layered with VSe$_2$.
The composition of the layer can differ from that of the (presumed) bulk structure
from which it is sliced.
The emergent Mn$_3$Se$_2$ structure cannot currently be identified as a slice of a
known bulk structure.
The misfit parameter $1+y$ equals the ratio of in-plane unit cell area of the
predicted structure to that of VSe$_2$, evaluated in the optimized island unit cell.
}
\label{tab:opt}
\end{center}
\end{table}

Table \ref{tab:opt} summarizes the systems predicted to form kinetically stable layers
between VSe$_2$ layers.
For each system, the table lists the composition found in the layer.
Most layers approximate a slice from a presumed bulk structure
with a sometimes different composition.
The structure type of the presumed bulk structure is given, along with
the orientation of the slice.
The last column lists the misfit parameters, calculated from the ratio of
the two constituents' approximate lattice parameters.

DFT calculations using
the {\sc VASP} package \cite{kresse96, kresse99}
serve to optimize initial structures and explore systems
that show promise.
The calculations allow spin polarization and
evaluate the energies and forces in
the generalized gradient approximation of
Perdew, Burke, and Ernzerhof \cite{PBE96}.
The electronic structure is treated with a Fermi-Dirac smearing of
width $0.1$ eV and converged to $10^{-5}$ eV.
The stopping criterion for ionic relaxation is $10^{-4}$ eV.
The calculations employ a single k-point ($\Gamma$).
The contributions from the
van der Waals interaction are described by the DFT-D2
pair-wise empirical approximation of Grimme
with the default values for the parameters \cite{grimme2006}.
The islands' initial structures are isostructural to truncated slices of binary crystal
structure types including auricupride, cadmium iodide,
iron stannide, molybdenum disilicide, and rock salt.
Typical unit cells contain between 18 and 70 ions in the island structure
and either 36 or 49 VSe$_2$ primitive cells in the continuous VSe$_2$ layer.
The initial structures are relaxed via the conjugate gradient method
with no imposed constraints.

For each of the three chemical combinations, the optimization of the initial
structures leads to a few
promising cases among many inauspicious cases.
The promising cases, detailed below, are characterized by intact VSe$_2$
layers and islands exhibiting a core region with well-defined structure.
The cases that show no promise in general do so very clearly:
the island structures often appear disordered,
and in some cases the VSe$_2$ layers disintegrate.

Figure \ref{fig:SSV} shows two optimized unit cells for VSe$_2$ layers confining a
SnSe$_2$ island, exemplifying a promising and an unfavorable case.
The promising SnSe$_2$ island (Fig.\ \ref{fig:SSV}(a)) stabilizes in the 1T structure,
approximately a triatomic thick (001) slice of SnSe$_2$
with the same cadmium iodide crystal structure type as VSe$_2$.
The layers differ in orientation by $26^\circ$.
The V-V bond lengths exhibit a narrow distribution
(with standard deviation $\sigma=0.02$~\AA) centered a{{round 3.28~\AA,
while the Sn-Sn bond lengths average 3.85~\AA\ with a much
wider $\sigma=0.26$~\AA.
Optimizing a unit cell with roles reversed
(i.e., complete SnSe$_2$ layer and VSe$_2$ island) leads to
a more accurate estimate of 3.82~\AA\ ($\sigma=0.02$~\AA)
for the Sn-Sn bond lengths.
These values result in a misfit parameter of 1.36, somewhat larger than the experimental
value of 1.25 \cite{kylepriv}.
Optimization of a SnSe-VSe$_2$ unit cell with a SnSe island based on the rock salt structure
(not shown here)
results in similar good agreement (see Ref.~\citenum{PhysRevB.91.144203})
with experiment \cite{Atkins2013128}.
The poorly structured island in Fig.\ \ref{fig:SSV}(b) represents one of many unfavorable
cases.
Among all the optimized unit cells that layer VSe$_2$ with a Sn-Se island, only two
qualify as promising cases, and these two show the same structure as the two
synthesized systems with this chemistry.

\begin{figure} %%%%%%%%%%%%%%%%%%%%%%%%%%%%%%%%%%%%%
\includegraphics[width=8.46cm]{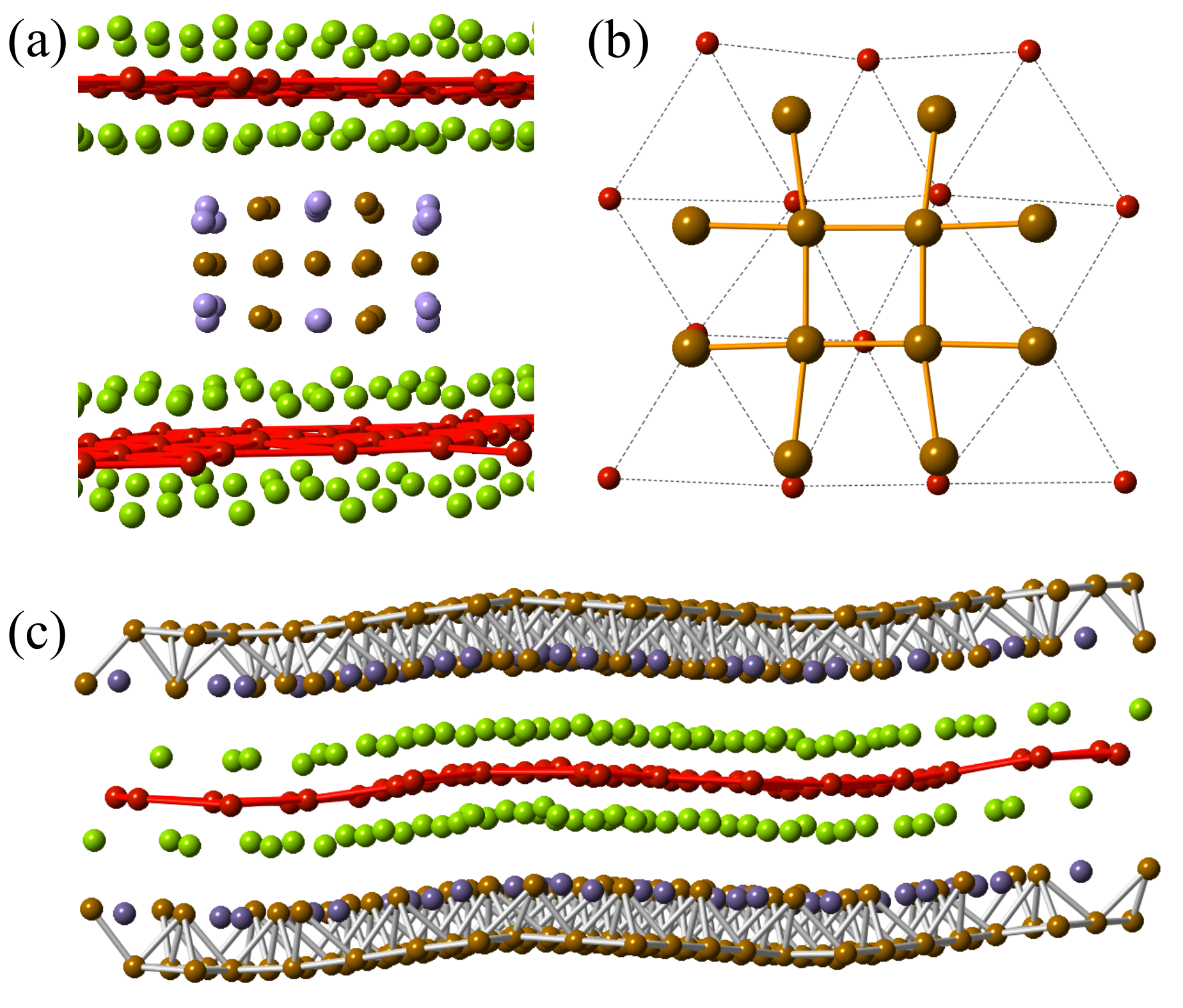}
\caption{\label{fig:FeBased}
Predicted layered Fe$_3$Ge-VSe$_2$ system.
Optimized island in
(a) side view and (b) top view showing orientation of Fe ions over V ions.
(c) Side view of optimized unit cell with complete VSe$_2$ and Fe$_3$Ge layers.
V, Se, Fe, and Ge ions are colored red, green, brown, and lavender, respectively.
}
\end{figure}

Having successfully reproduced the known structures that interleave VSe$_2$ with
Sn-Se layers, the approach is used to predict systems that have yet to be synthesized.
Figure \ref{fig:FeBased} shows the predicted Fe$_3$Ge-VSe$_2$ system.
The Fe$_3$Ge constituent is a triatomic thick (001) slice of bulk cubic
$\varepsilon_1$-Fe$_3$Ge (AuCu$_3$ structure),
exhibiting bond lengths close to that of the bulk material.
The Fe$_3$Ge island has average Fe-Fe bond lengths of 2.48~\AA\ 
($\sigma= 0.08$~\AA)
and average V-V bond lengths of 3.34~\AA\ ($\sigma= 0.19$~\AA).
An optimized unit cell with complete Fe$_3$Ge layers
and VSe$_2$ islands
has average Fe-Fe bond lengths of 2.55~\AA\ 
($\sigma=  0.08$~\AA),
closer to
the measured Fe-Fe bond length of 2.59~\AA\ in bulk $\varepsilon_1$-Fe$_3$Ge
\cite{kanematso64}.

Optimized island-containing unit cells of the promising cases
generally have parallel, planar layers.
Approximate unit cells with both layers completed often
show a wave-like structure as in 
Fig.\ \ref{fig:FeBased}(c).
The spatial oscillations are commensurate with the unit cell
and arise from strain imposed on one or both layers.
For example,
the unit cell in Fig.\ \ref{fig:FeBased}(c) contains a VSe$_2$ layer with in-plane lattice vectors 
that are around 3\% larger than those of bulk VSe$_2$,
while the in-plane lattice vectors of the Fe$_3$Ge layer
are around 5\% shorter than those of bulk Fe$_3$Ge.
With both layers strained, both show the spatial oscillation.

\begin{figure} %%%%%%%%%%%%%%%%%%%%%%%%%%%%%%%%%%%%%
\includegraphics[width=8.46cm]{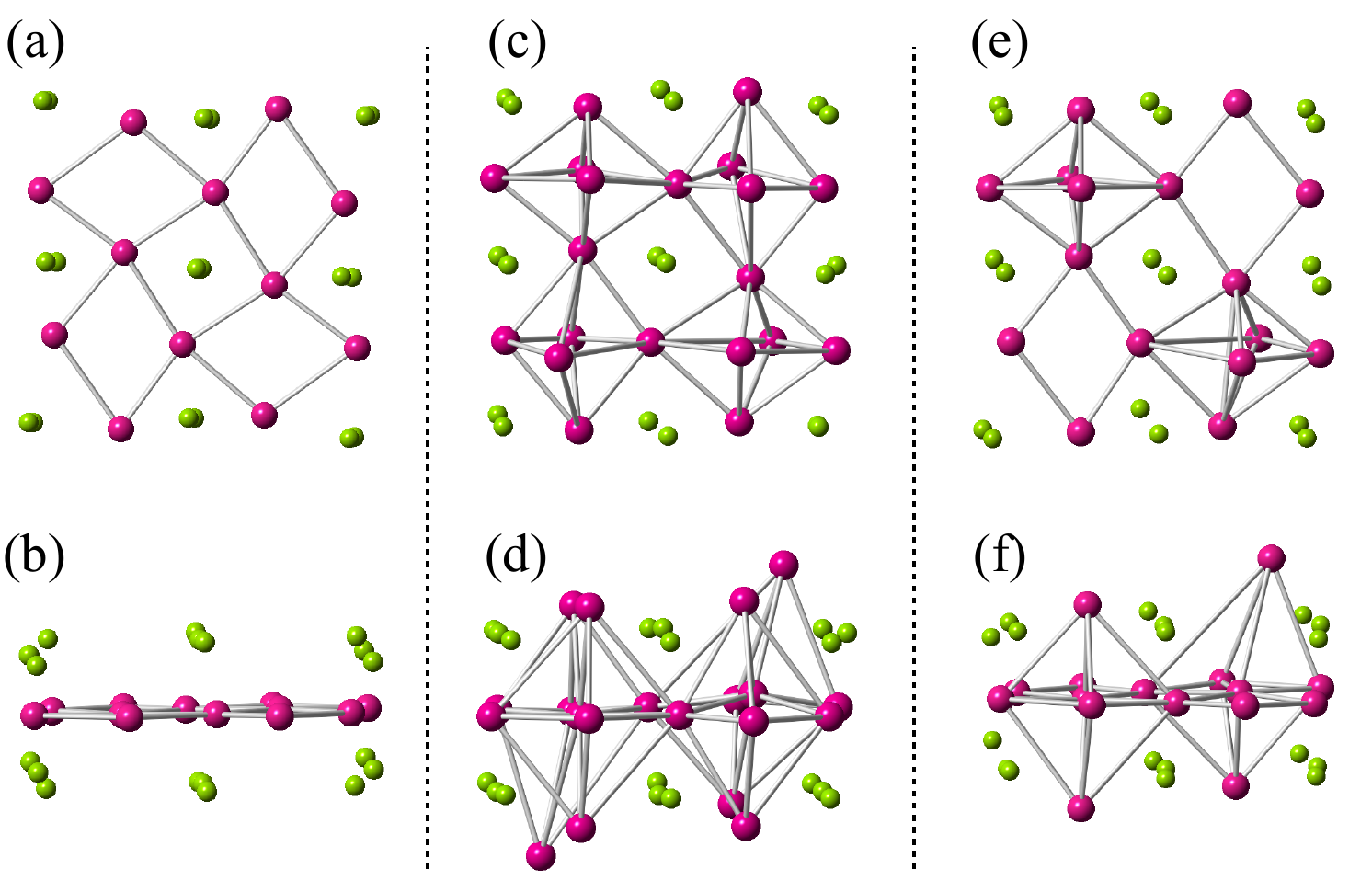}
\caption{\label{fig:MSV3}
Top and side views of optimized islands (VSe$_2$ layers not shown)
for
(a, b) MnSe-VSe$_2$,
(c, d) Mn$_2$Se-VSe$_2$, and
(e, f) Mn$_3$Se$_2$-VSe$_2$.
Se and Mn ions are colored green and magenta, respectively.
}
\end{figure}

The character of the spatial oscillation of the unit cell in Fig.\ \ref{fig:FeBased}(c) 
re-emerges in the magnetic structure.
Bader analysis of the up and down spin charge distributions serves to assign
their sum and difference to charge and magnetic moment,
respectively, of the individual ions \cite{Henkelman2006354}.
The magnetic moments on the V ions lie between 0 and 1.6 $\mu_\text{B}$
and exhibit a spin density
wave commensurate with the wave-like structure shown in Fig.\ \ref{fig:FeBased}(c).
The same pattern appears in the magnetic moments of the Fe ions, which range between 1.5 and
2.2 $\mu_\text{B}$.
The Bader analysis for the charge distribution suggests a small charge transfer from the
Fe$_3$Ge layer to the VSe$_2$ layer of approximately 0.2 electrons per Fe ion.

The search for systems interleaving VSe$_2$ and Mn$_x$Se$_y$ layers
uncovers multiple stable islands,
two with closely related structures.
Figures \ref{fig:MSV3}(a) and (b) show the first island, a finite cut from
a triatomic thick (001) slice of MnSe in a form isostructural with iron stannide.
The lattice parameter values average to 3.76~\AA.
In an optimized unit cell with both MnSe and VSe$_2$ layers completed,
the MnSe layer contains octahedra
with Se ions at the apexes and Mn ions forming the central-layer square.
These octahedra
exhibit a checkerboard rotational pattern
with the octahedra rotated by approximately $\pm 7 ^\circ$.

Figures \ref{fig:MSV3}(c) and (d) show the second island containing Mn and Se ions.
The Mn$_2$Se layer
resembles a triatomic thick (001) slice of Mn$_3$Se in a
distorted AuCu$_3$ structure.
The lattice parameter values average to 3.58~\AA.
This structure resembles the previous MnSe island, but with additional Mn ions
centered above the Mn diamond shapes of the central monolayer.
This reduces the rotation of the MnSe octahedra by approximately 2$^\circ$.
The additional Mn ions sit slightly out of the monolayer formed by the Se ions
and do so with unequal amounts, accompanied by a slight in-plane displacement
away from high-symmetry positions.

The two independently optimized 
MnSe and Mn$_2$Se
islands differ essentially only in the eight Mn ions,
suggesting possible intermediate systems with less than all eight sites occupied.
The optimization of such systems, constructed by selectively removing different
numbers of the eight Mn ions, results in stable islands.
Comparison of the energies reveals that the most stable system 
is Mn$_3$Se$_2$-VSe$_2$, which
contains four of the eight
Mn ions in the pattern shown in Fig.\ \ref{fig:MSV3}(e) and (f).
The lattice parameter averages 3.61~\AA\ (i.e., between the averages
of the MnSe and Mn$_3$Se islands).
The formation energies of the islands with between zero and eight Mn ions differ at
a scale an order of magnitude smaller than room temperature, suggesting a system
with potentially interesting properties related to variable composition, rotation of the SeMn octahedra,
disorder among occupied Mn sites, and magnetic structure.

In addition to the stable islands shown in Fig.\ \ref{fig:MSV3},
two other Mn$_x$Se$_y$ systems optimize into stable island structures
interleaved with VSe$_2$.
The first is isostructural with SnSe$_2$-VSe$_2$.
The MnSe$_2$ layer optimizes with slightly smaller lattice constant $3.65$~\AA,
giving this system a slightly smaller misfit parameter of 1.24.
The second system stabilizes with a two monolayer MnSe$_2$ structure that
resembles a section of the molybdenum disilicide structure type.
The layer is a $[100]$ slice of the crystal structure with in plane lattice parameters
of approximately 3.1~\AA\ and 7.6~\AA.
The unit cell outlined by these lattice parameters contain two MnSe$_2$ formula units,
thereby doubling this system's misfit parameter compared to
the other systems' typical values.

Calculated formation energies $E_\text{f}$ for the five Mn$_x$Se$_y$-VSe$_2$ systems
provide an estimate for their stability.
The values for $E_\text{f}$ are calculated for approximate unit cells with completed layers,
with the energies for VSe$_2$, Mn, and Se in their equilibrium bulk form serving as
references.
The $E_\text{f}$ are at best approximate, as these calculations use
a k-point mesh with density $20/$\AA$^{-1}$
(computational resources prevent using a finer grid).
MnSe$_2$ in the cadmium iodide structure and MnSe in the iron stannide structure
appear most stable with a predicted $E_\text{f}$ of approximately -0.1 eV/ion.
The Mn$_3$Se$_2$ structure has an estimated formation energy of roughly -0.05 eV/ion,
and hence lies above the convex hull.
MnSe$_2$ in the auricupride structure requires roughly +0.02 eV/ion to form.
No unit cell is constructed for the MnSe$_2$ system in the molybdenum disilicide
structure; the formation energy for this system in the island approximation is
approximately +0.07 eV/ion.

The predicted Mn$_x$Se$_y$-VSe$_2$ systems exhibit either
slices of known Mn$_x$Se$_y$ bulk crystal structures or
form unrelated structures that emerge in the nanoscale confinement.
Either option could be expected, as 
none of the known bulk Mn$_x$Se$_y$ crystal structures exhibit a 2D character \cite{Jain2013}.
Among the known bulk crystal structures with MnSe composition,
the high-pressure nickeline structure \cite{cemic1972}
and the rock salt structure \cite{cook1968}
can both be sliced
into a three monolayer piece that approximates a slice of the cadmium iodide
structure type.
The known bulk MnSe$_2$ crystal structure \cite{elliott1937}, of pyrite structure type,
resembles none of the stable island structures.
However, the pyrite structure type shares one structural component with the
molybdenum disilicide structure type MnSe$_2$ layer:
the MnSe$_2$ layer exhibits parallel lines that repeat a Se-Mn-Se pattern;
the MnSe$_2$ crystal structure contains the same Se-Mn-Se patterned lines,
but instead of aligning parallel the lines lie at oblique angles.

In conclusion, the theoretical approach outlined here proposes to accelerate
the experimental discovery of new layered materials
as well as the theoretical search for promising systems.
It layers two constituents of nanoscale thickness:
one from an existing layered material (here VSe$_2$),
the second from sets of chemical elements (here period four elements)
to probe for stable structures.
The probing relies on density functional theory calculations to optimize
initial structures,
with the second layer approximated as a finite island.
This approximation dramatically accelerates the theoretical search for
new materials because (a) the unit cells contain significantly fewer ions 
and (b) a single initial structure serves as starting point for multiple
final candidate structures.
The approach results in predictions of kinetically stable layered materials
as guidance for synthesis efforts.
The results outlined here specify systems amenable to synthesis: VSe$_2$
interleaved with either Fe$_3$Ge or Mn$_x$Se$_y$.
In particular,  Mn$_x$Se$_y$-VSe$_2$ promises interesting behavior
including emergent structures,
and preliminary synthesis attempts of Mn$_x$Se$_y$ layered with VSe$_2$
show promise \cite{kylepriv}.

This research is supported by the Department of Energy under
Contract No.\ DE-AC52-06NA25396
and Grant No.\ LDRD-DR 20140025.
The author thanks
E. Chisolm,
D. Hamann,
R. Hennig,
O. Hite,
D. Johnson,
A. Niklasson,
B. Revard,
and
F. Ronning
for helpful and encouraging discussions.

\providecommand{\latin}[1]{#1}
\providecommand*\mcitethebibliography{\thebibliography}
\csname @ifundefined\endcsname{endmcitethebibliography}
  {\let\endmcitethebibliography\endthebibliography}{}

\end{document}